\documentclass{article}
\usepackage{waspaa23,amsmath,graphicx,url,times}
\usepackage{color}
\usepackage{hyperref}
\usepackage{amsfonts, amssymb}
\usepackage{url}
\usepackage{cite}
\usepackage{hyperref}
\usepackage{array,eqparbox,collcell}
\usepackage{gensymb}
\usepackage{xcolor}
\usepackage{balance}

\title{Mitigating Cross-Database Differences for Learning Unified HRTF Representation}

\name{Yutong Wen, You Zhang, Zhiyao Duan \thanks{This work is partially supported by the Goergen Institute for Data Science at the University of Rochester, a New York State Center of Excellence in Data Science award, and synergistic activities funded by the National Science Foundation (NSF) under grant DGE-1922591. 
}}
\address{University of Rochester, Rochester, NY, USA
}

\begin{document}

\ninept
\maketitle

\begin{sloppy}

\begin{abstract}
  Individualized head-related transfer functions (HRTFs) are crucial for accurate sound positioning in virtual auditory displays. As the acoustic measurement of HRTFs is resource-intensive, predicting individualized HRTFs using machine learning models is a promising approach at scale. Training such models require a unified HRTF representation across multiple databases to utilize their respectively limited samples. However, in addition to differences on the spatial sampling locations, recent studies have shown that, even for the common location, HRTFs across databases manifest consistent differences that make it trivial to tell which databases they come from. This poses a significant challenge for learning a unified HRTF representation across databases.
  In this work, we first identify the possible causes of these cross-database differences, attributing them to variations in the measurement setup. Then, we propose a novel approach to normalize the frequency responses of HRTFs across databases. We show that HRTFs from different databases cannot be classified by their database after normalization. We further show that these normalized HRTFs can be used to learn a more unified HRTF representation across databases than the prior art. We believe that this normalization approach paves the road to many data-intensive tasks on HRTF modeling. 
\end{abstract}


\section{Introduction}
\label{sec:intro}

Head-related transfer functions (HRTFs) represent the spectral changes that occur as a sound wave travels from a source to the listener's ears, taking into account the listener's unique ear, head, and torso geometry. They serve as a crucial cue for human auditory systems to localize sound sources in a 3D space. HRTFs are fundamental in virtual auditory displays, where they are used to create 
more accurate localization in virtual environments~\cite{Xie2014HRTF}. 

HRTFs are different from person to person due to the geometry differences. To achieve natural auditory displays, individualized HRTFs are preferred over generic ones since the latter often cause 
front-back confusions, distance errors, and other related issues, particularly at locations with a lack of interaural differences~\cite{møller1996binaural}. 
The acoustic measurement method, which is the conventional approach for obtaining individualized HRTFs, is resource-intensive in terms of both time and equipment. A typical measurement setup, situated within an anechoic room, includes a vertical array of loudspeakers arranged around an arc and a robotic arm for controlling the horizontal rotation of the arc~\cite{bili}. Several research labs have conducted measurements to obtain individualized HRTFs and published HRTF databases~\cite{3d3a, bili,cipic,hutubs,ita,riec,sadie}, where each of them contains dozens of subjects.


In recent years, researchers have been developing machine learning methods to estimate individualized HRTFs from geometry information~\cite{ZIEGELWANGER2016threeDmesh, chun2017deep, Lee2018P-HRTF, wang2021global, wang2022predicting}.
These methods are trained and evaluated on a single database, which usually comprises only several dozen subjects, thereby limiting models' accuracy and generality to a larger population. A natural strategy to improve the performance of these methods is to enlarge the training set by combining multiple HRTF databases. 
However, different databases often adopt different spatial sampling schemes, and   
mix-database training requires a unified representation across different spatial sampling schemes. Such representation has been attempted by Zhang et al.~\cite{Zhang2022HRTFfield}, where \textit{HRTF field} was proposed as a neural field representation learned from different HRTF databases with different spatial sampling schemes. 



In addition to different spatial sampling schemes, a recent study by Pauwels et al.~\cite{pauwels2022relevance} has revealed that there are other significant discrepancies across HRTF databases; these discrepancies are so significant that HRTFs from different databases from the same source locations can be easily classified by a support vector machine (SVM) classifier.
This finding suggests that there are systematic differences beyond spatial sampling schemes in HRTF data across databases, and these differences are likely to hinder the training of more accurate and generalizable machine learning models for HRTFs on merged databases. 

In our work, we start by analyzing the potential factors for systematic cross-database differences and illustrate position-dependent differences on average-person HRTF frequency responses. 
Then, we propose a novel normalization strategy to reduce the cross-database differences in HRTF frequency responses; We show that this normalization strategy makes HRTFs from different databases hard to distinguish by a kernel support vector machine (SVM) classifier. Finally, we apply the normalization strategy to \textit{HRTF field}\cite{Zhang2022HRTFfield}, a state-of-the-art method for learning a unified HRTF representation across databases; We show that the proposed normalization strategy improves the unified representation learning, resulting in prominent improvement on HRTF reconstruction accuracy on a cross-database reconstruction task.
Our code is available at: \url{https://github.com/YutongWen/HRTF_field_norm}.

\begin{table*}[t]
\centering
\begin{tabular}{c|cccccccccc}
\hline
Database Information & ARI  & ITA   & Listen & Crossmod & SADIE II & BiLi & HUTUBS & CIPIC & 3D3A & RIEC \\ \hline
\# Subjects         & 97   & 48    & 50     & 24       & 18       & 52   & 96     & 45    & 38   & 105  \\
\# Positions        & 1550 & 2304* & 187    & 651      & 2818*    & 1680 & 440    & 1250  & 648  & 865  \\
Source Distance (m)  & 1.2  & 1.2   & 1.95   & 1.0      & 1.2      & 2.06 & 1.47   & 1.0   & 0.76 & 1.5  \\ \hline
\end{tabular}
\caption{Database information. Position number with an asterisk * signifies that there are variations for some subjects in that database.} 
\label{tab:dataset_info}
\end{table*}


\section{Investigating cross-database differences 
}
\label{sec:dataset_characteristics}





HRTFs describe the spatial filtering effect of the listener's head, torso, and outer ears on sound arriving from different directions. However, measured HRTFs often contain effects from the measurement equipment and recording environment as well, which may be the main contributors to the cross-database differences discovered in~\cite{pauwels2022relevance}. We present ten selected databases: 3D3A~\cite{3d3a}, ARI, BiLi~\cite{bili}, CIPIC~\cite{cipic},  Crossmod, HUTUBS~\cite{hutubs}, ITA~\cite{ita}, Listen, RIEC~\cite{riec}, SADIE II~\cite{sadie},       and their information is shown in Table~\ref{tab:dataset_info}.


\begin{figure}[t]
  \centering
  \includegraphics[width=0.99\linewidth]{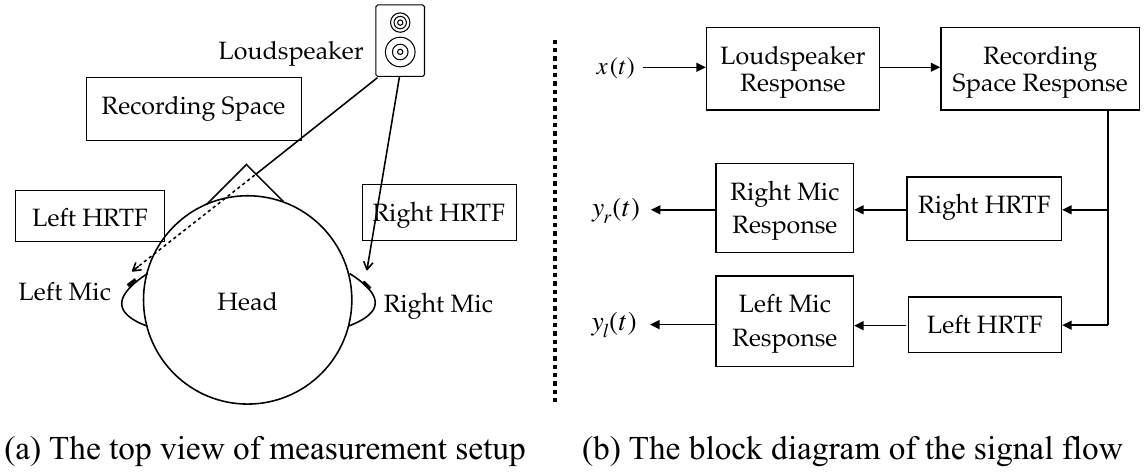}
  \caption{Illustration of HRTF measurement.} 
  \label{fig:signal_flow}
\end{figure}

Figure~\ref{fig:signal_flow} illustrates the top view of a general measure setup and the signal flow from the sound source to in-ear microphones. At a particular location denoted by azimuth $\theta$ and elevation $\phi$, a source signal $x(t)$ is played back by a loudspeaker. The resulting sound wave propagates through the recording environment (typically an anechoic chamber) and reaches the ear canals of a listener. The sound wave is then recorded by the in-ear microphones as $y_l(t)$ and $y_r(t)$. HRTFs are then computed by deconvolving the received signals $y_l(t)$ and $y_r(t)$ with the source signal $x(t)$, assuming that the signal path is a linear time-invariant system.

One systematic difference across HRTF databases is caused by the frequency response of the loudspeaker. While high-quality loudspeakers are often used in HRTF measurement, the frequency responses of different loudspeakers can be quite different. In addition, loudspeakers used in HRTF measurement are directional loudspeakers, meaning that their frequency responses vary with the direction between the receiver and the loudspeakers. 
This variation can be significant when the loudspeaker is close to the listener. 

Another difference may arise from the frequency responses of microphones. Different microphones used in different databases are likely to have different frequency responses. In addition, different microphones are placed at different depths into the ear canal, causing further variations in frequency responses.

Regarding the recording space, while it is typically intended to be anechoic, the reflections may still cause changes to frequency response
due to the geometry and material of measurement equipment (e.g., mechanic arm, soundproofing foam, chair, and floor) that makes the space not entirely anechoic. This response may also be dependent on the source location.

\begin{figure}[t]
  \centering
  \includegraphics[width=0.98\linewidth]{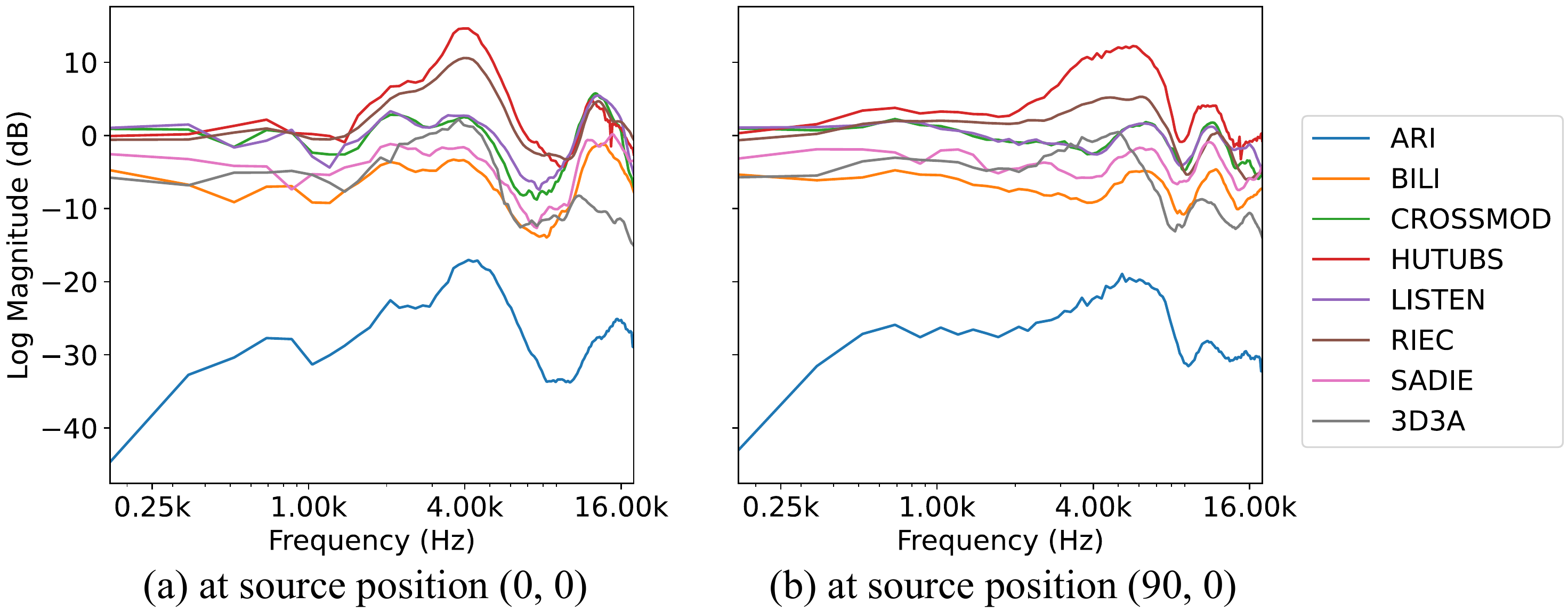}
  \caption{Average person HRTFs at locations (0, 0) and (90, 0) from eight measured far-field databases.} 
  \label{fig:system_response}
\end{figure}

To empirically investigate the differences in system responses across different databases, we conducted an analysis of the average HRTFs across individuals within each database at specific source positions. We pick the common locations across all databases to remove the effects of the different spatial schemes. To maximize the usage of data and reduce the variance of average, we then map the right ear HRTF at location $(\theta, \phi)$ to its corresponding left ear HRTF at location $(2\pi - \theta, \phi)$. This is to assume the symmetry between the left and the right ear HRTFs of each ``augmented'' subject. 
This process allows us to double our sample size for the left ear. As such, the following discussions of HRTFs are all about the left ears. 
This mapping strategy has also been applied to the other experiments in this study. %


\begin{figure*}[t]
  \centering
  \includegraphics[width=0.99\linewidth]{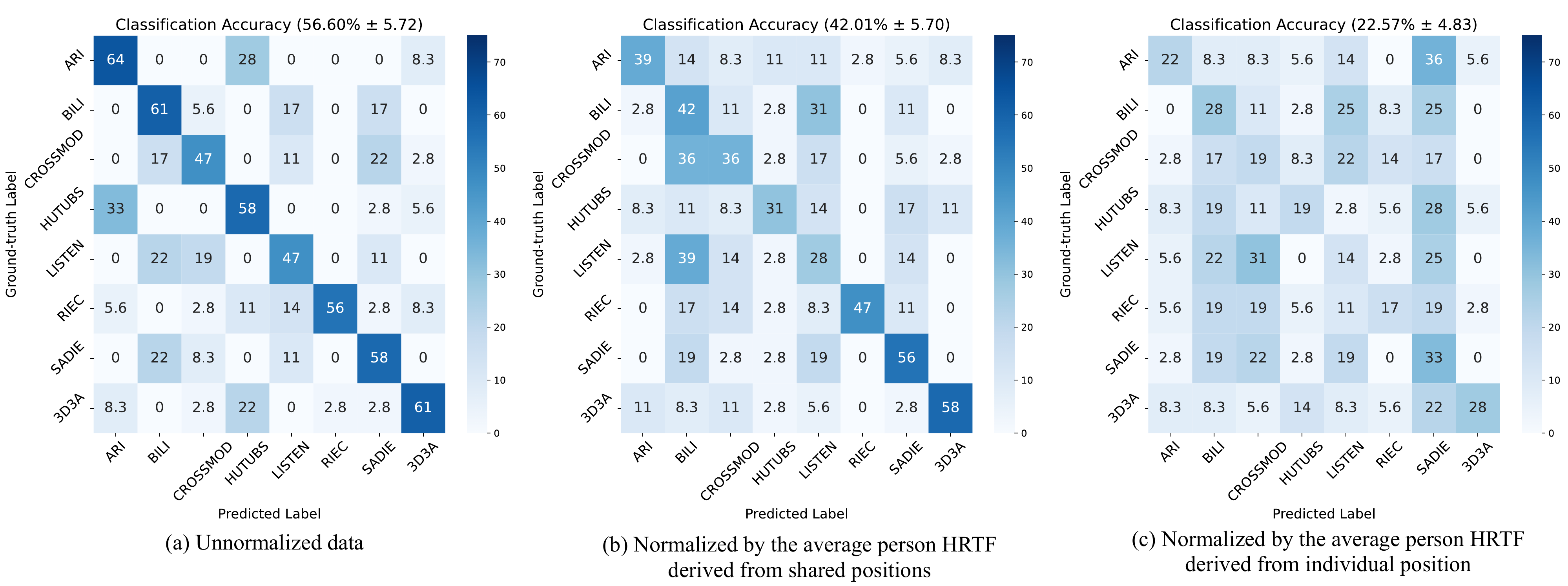}
  \caption{Confusion matrices of HRTF database classification with kernel SVM on original HRTFs and HRTFs normalized with two strategies.} 
  \label{fig:cm_results}
\end{figure*}



Figure~\ref{fig:system_response} shows the average HRTF frequency responses of eight different databases at two common source locations: front (0,0) and right (90,0). A frequency range from 200Hz to 18kHz is displayed in the log-frequency scale.
We can see that different databases show discernible discrepancies in the average HRTF frequency responses at both source positions, suggesting systematic differences in the measurement system responses in these databases. In addition, the difference in the system responses across databases does not remain unchanged when the source position varies, suggesting that the system response is position dependent. We argue that these position-dependent systematic differences in measurement system responses need to be removed or reduced if one would like to combine different HRTF databases to train machine learning models. In the next section, we will propose a strategy to reduce these differences. 

\section{Mitigating the Spectral Discrepancy across Different Databases}
\label{sec:mitigate_discrepancy}
Based on the findings in the previous section, spectral variations exist among databases and across different source positions, which constitute potential factors impairing the generalization and accuracy of machine learning models. In this section, we propose a normalization approach motivated by the previous section to address systematic spectral differences across different databases. The effectiveness of our method is evaluated by comparing the performance of a kernel SVM classifier trained on the normalized data against that of the original (unprocessed) data.





For a given subject's HRTF $Y(\theta, \phi)$ measured from the source position $\theta$ and $\phi$ in compliance with the database's average person HRTF measured from the same source position $\textit{HRTF}_\mathrm{avg}(\theta, \phi)$, we propose to normalize its HRTF by 
\begin{equation}
    \textit{HRTF}_{\mathrm{normalized}}(\theta, \phi) = \frac{Y(\theta, \phi)}{\textit{HRTF}_{\mathrm{avg}}(\theta, \phi)},
\end{equation} 
where $\textit{HRTF}_{\mathrm{normalized}}(\theta, \phi)$ denotes the normalized HRTF.

To assess the effectiveness of our normalization results, we trained a kernel support vector machine (SVM) following \cite{pauwels2022relevance} to classify which database a given HRTF originates from. This particular selection was based on the consideration that kernel SVM classifiers are less susceptible to overfitting in comparison to complex models such as multilayer perceptrons with limited training data.

Following the methodology described in~\cite{pauwels2022relevance},
our study proceeded to train the kernel SVM on the subset of twelve positions which were found to be shared across eight of the selected databases; the remaining two databases do not contain these positions in common. In this task, a total of 144 subjects were utilized, with 18 subjects (the smallest database size) extracted from each database. This sampling methodology was adopted to reduce bias toward any database. The total number of HRTFs in this experiment is thus 12 (positions) $\times$ 18 (subjects) $\times$ 2 (ears) = 432 for each database.
We perform five-fold cross-validation and report the mean accuracy and confusion matrix in Figure~\ref{fig:cm_results}.

First, we present the SVM classification results on the unnormalized data shown in Figure~\ref{fig:cm_results}(a) as the baseline. The results are close to what is reported in~\cite{pauwels2022relevance} as a sanity check. The classification accuracy is a little lower than their reported 61.2\%. This is because we discarded the two simulated databases they used, which achieved much higher ($>$90\%) accuracy and hence increased their overall classification accuracy. Then, in Figure~\ref{fig:cm_results}(b), we provide the outcomes obtained from position-independent normalization, computed by normalizing the mean average-person HRTFs across 12 shared positions. Finally, we show the results derived from our proposed normalization method in Figure~\ref{fig:cm_results}(c), which is conducted on an individual location and per ear basis.

Comparing the results depicted in Figure~\ref{fig:cm_results} (b) (c) to (a), both accuracy from normalizing by averaging shared positions and by individual position is lower than the accuracy from the unnormalized data, suggesting that both normalization methods are able to confuse the classifier, hence effective in reducing systematic differences across databases. Further, comparing the results in Figure~\ref{fig:cm_results} (b) to (c), position-dependent normalization accuracy is much lower than the position-independent normalization accuracy, reinforcing our analysis in Section~\ref{sec:dataset_characteristics} that the systematic cross-database differences are more likely to be dependent on source position. 
Note that differences due to the overall normalization of the datasets are ruled out as a cause.
The normalization technique applied to individual locations, with results shown in Figure~\ref{fig:cm_results}(c) yields very low classification accuracy, where the accuracy is close to a random guess of classifying into one of the databases. This suggests that the cross-database differences are highly reduced to the extent that the SVM classifier becomes confused. This shows the success of our proposed normalization on an individual position basis.

\section{Mix-database training with neural fields}
\label{sec:mix_data_training}

\begin{figure}[t]
  \centering
  \includegraphics[width=\linewidth]{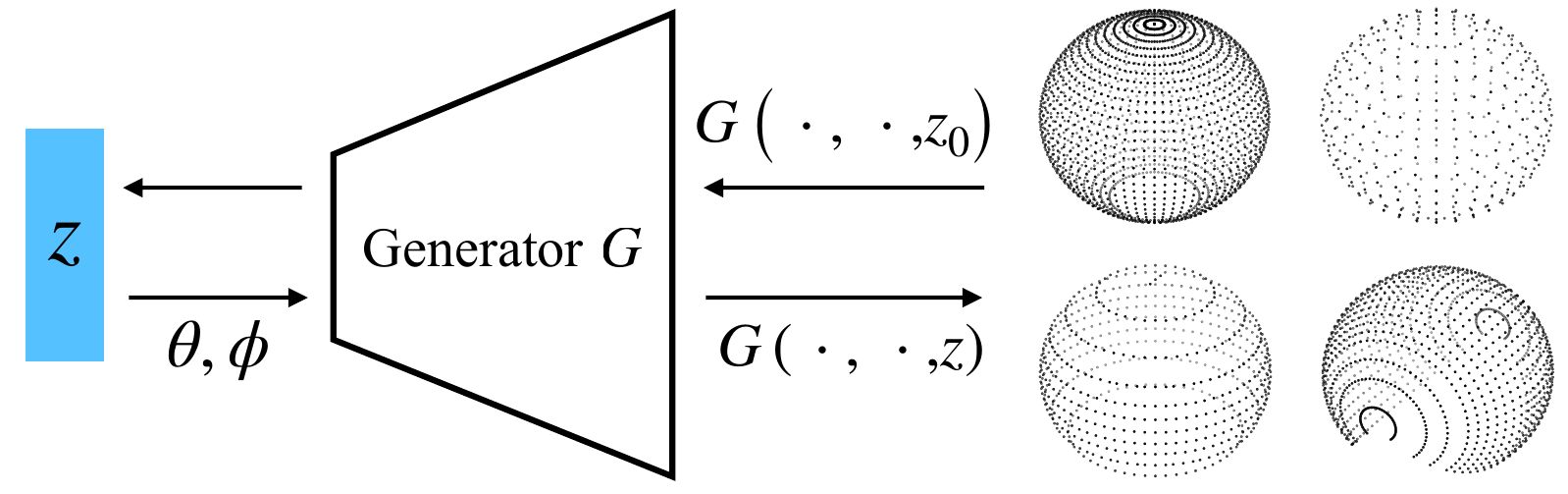}
  \caption{Unified HRTF representation learning with neural fields~\cite{Zhang2022HRTFfield} on mix databases with proposed normalization strategy.}
  \label{fig:mix_training}
\end{figure}


To further evaluate the benefits of our proposed normalization method, we apply it to training machine learning models for learning a unified HRTF representation across databases.


\subsection{Experimental setup}

\textbf{Model}: \textit{HRTF field}~\cite{Zhang2022HRTFfield} has shown successful mix-databases training since they learned a unified representation across different spatial sampling schemes used by different databases. 
The model represents each person's HRTF with $G(\theta, \phi, z)$, where $z$ represents one person's identity. The training and evaluation both involve two steps, as illustrated in Figure~\ref{fig:mix_training}. The first step is to infer a latent vector $z$ from the gradient of the mean squared error (MSE) between the ground truth and setting $z$ at $z_0$. The second step is to update the generator with the MSE loss. After convergence, the generator would learn a unified HRTF representation.

\textbf{Task}: Instead of the interpolation and the generation tasks in the original paper~\cite{Zhang2022HRTFfield}, we propose cross-database reconstruction since we focus on mitigating the cross-database differences. We train the generator with multiple databases combined. During inference, for a given HRTF from another database (different from databases used in training), we employ the trained model to infer $z$ and reconstruct this HRTF. 

\textbf{Evaluation}: The log-spectral distortion (LSD) is utilized as the evaluation metric, and it is defined as
\begin{equation}
\textit{LSD}(H, H') = \sqrt{\frac{1}{PN}\sum_{\theta, \phi}\sum_{n}\left (20 \log_{10}\left | \frac{H(\theta, \phi, n)}{H'(\theta, \phi, n)}  \right | \right  )^2  },
\label{eq: lsd}
\end{equation}
where $H(\theta, \phi, n)$ and $H'(\theta, \phi, n)$ denote the $n$-th frequency bin ($N$ in total) of the ground-truth and predicted HRTFs measured at the source position $(\theta, \phi)$, respectively. $P$ represents the number of total source positions within the database.

As the model is trained on normalized HRTFs, the predicted HRTFs need to be de-normalized using the same normalization factors before computing the LSD with the ground truth. 
However, as the normalization is a division by the average-person HRTF and LSD is in the dB scale, LSD can also be equivalently computed between the predicted HRTFs and the normalized ground truth.

\textbf{Baseline}: In the original paper of HRTF field~\cite{Zhang2022HRTFfield}, the authors applied normalization of the equator energy, but we argue that it is insufficient as it is not frequency dependent nor position dependent. In the following sections, we will compare the performance of our proposed normalization techniques with the original ones.




\subsection{Cross-database reconstruction performance of different normalization methods}
To more effectively demonstrate our improvement, we intentionally construct 5 experiments to incorporate different cases and present the comparison between the obtained results using our normalization method with the baseline in Table \ref{tab:lsd_table}.

In order to evaluate the generalization capability of our proposed method, we first select several databases (BiLi, Crossmod, Listen, SADIE II) that exhibit similar system frequency responses in Figure~\ref{fig:system_response} for training, and a database (ARI) with very different system frequency responses for testing.
This would pose a significant challenge to the model's generality. The results are shown under experiment 1 in Table~\ref{tab:lsd_table}. The results show that our proposed approach outperformed the baseline LSD of 7.47 dB, achieving a markedly lower LSD value of 4.69 dB. These results provide empirical support for the effectiveness of our proposed methodology in enhancing the model's generality.

The second experiment is designed to assess the model's capacity to cope with varying system frequency responses in training and testing databases. The databases are selected to ensure that no two databases exhibit similar system frequency responses. Results are shown under experiment 2 in Table~\ref{tab:lsd_table}, showing that our proposed methodology effectively mitigated the model's challenges in dealing with distinct system frequency responses, yielding a notable improvement over the baseline LSD of 5.54 dB, with an LSD of 4.82 dB. This underscores the potential of our approach to enhance the model's capacity to handle variability in system frequency responses across databases.

The third experiment is devised such that
the training and testing databases manifest highly similar system frequency responses. The result confirmed the effectiveness of our proposed method in enhancing the model's performance,
with the LSD improving from 4.31 dB to 3.89 dB. This provides compelling evidence of the versatility of our proposed approach, and its capacity to enhance the model's performance across different scenarios. In the next two experiments (experiments 4 and 5 in Table~\ref{tab:lsd_table}), we use one database randomly chosen for evaluation and the other nine for training. In addition, we do 10-fold cross-validation on this setup, and the average LSD for our result is 4.24 dB and for the baseline is 4.83 dB.

\begin{table}[t]
\centering
\begin{tabular}{c|c|c|c|c|c}
\hline
    Experiments       & 1                & 2                & 3                & 4                & 5                \\ \hline
ARI        & $\bigcirc$       & $\bigtriangleup$ &                  & $\bigtriangleup$ & $\bigtriangleup$ \\
ITA        &                  &                  &                  & $\bigtriangleup$ & $\bigtriangleup$ \\
Listen     & $\bigtriangleup$ &                  & $\bigcirc$       & $\bigtriangleup$ & $\bigtriangleup$ \\
Crossmod                                                                   & $\bigtriangleup$ & $\bigtriangleup$ & $\bigtriangleup$ & $\bigtriangleup$ & $\bigtriangleup$ \\
SADIE II   & $\bigtriangleup$ &                  & $\bigtriangleup$ & $\bigtriangleup$ & $\bigtriangleup$ \\
BiLi                                                                       & $\bigtriangleup$ & $\bigtriangleup$ & $\bigtriangleup$ & $\bigtriangleup$ & $\bigtriangleup$ \\
HUTUBS     &                  & $\bigtriangleup$ &                  & $\bigtriangleup$ & $\bigcirc$       \\
CIPIC      &                  &                  &                  & $\bigtriangleup$ & $\bigtriangleup$ \\
3D3A       &                  &                  &                  & $\bigtriangleup$ & $\bigtriangleup$ \\
RIEC       &                  & $\bigcirc$       &                  & $\bigcirc$       & $\bigtriangleup$ \\ \hline
HRTF field~\cite{Zhang2022HRTFfield} & 7.47            & 5.54             & 4.31            & 4.43            & 5.01            \\
\textbf{Our proposed}       & \textbf{4.69}   & \textbf{4.82}   & \textbf{3.89}   & \textbf{3.73}   & \textbf{4.04}   \\ 
\hline
w/o position dependency & 5.61            & 5.32            & 4.32            & 4.00            & 4.89            \\
w/o ear dependency & 5.11             & 5.11            & 3.98            & 3.94            & 4.67            \\
\hline
\end{tabular}
\caption{Log-spectral distortion (LSD) in dB scale of cross-database HRTF reconstruction with five different experiments (columns) using different mix-databases. For each experiment, training sets are denoted with $\bigtriangleup$, evaluation sets are denoted with $\bigcirc$, and databases not used are left as blank.}
\label{tab:lsd_table}
\end{table}

\textbf{Ablation study}: We perform a comprehensive analysis of the significance of each step in our proposed methodology. This was accomplished through a comparison of results obtained from HRTF field \cite{Zhang2022HRTFfield} trained on mix-databases using common position normalization, common microphone normalization (wherein the left and right mics are treated as having identical frequency responses), and our proposed normalization approach. The experimental findings are presented in Table \ref{tab:lsd_table}.
As shown in the table, our proposed normalization method has the best result in each experimental setup, suggesting that each step in our normalization method is necessary. 


\section{Conclusion}
\label{sec:conclusion}
In this study, we identified the different position-dependent system frequency responses as a key factor causing the systematic differences across databases. We proposed a novel normalization strategy that effectively mitigates the influence of this factor. By employing our normalization method, we effectively confused the kernel SVM classifier in performing the task of distinguishing which database a given HRTF originates from. Additionally, we showed that our method improved the unified HRTF representation learning across different HRTF databases, resulting in notable improvement on HRTF reconstruction accuracy. 
We believe our method represents a promising step towards integrating existing HRTF databases and can benefit future machine learning approaches to tackle challenges on data-intensive HRTF modeling. Overall, our findings contribute to a better understanding of the factors affecting HRTF database misalignment and provide a practical solution for enhancing the accuracy and generalization of HRTF representation learning. Future work includes HRTF normalization and de-normalization strategies for novel source positions.

\balance
\bibliographystyle{IEEEtran}
\bibliography{main}

%
%
%
%
%
%
%
%
%

\end{sloppy}
\end{document}